\documentclass[aps,prl,amsmath,amssymb,reprint,superscriptaddress,]{revtex4-2}
\usepackage{charter,graphicx,verbatim,threeparttable,float,amssymb }
\usepackage[version=4]{mhchem}

\begin{document}

\title[KV$_3$Sb$_{5-x}$Sn$_x$ and RbV$_3$Sb$_{5-x}$Sn$_x$]{Tuning charge-density wave order and 
superconductivity in the kagome metals KV$_3$Sb$_{5-x}$Sn$_x$ and RbV$_3$Sb$_{5-x}$Sn$_x$}

\author{Yuzki M. Oey}
\affiliation{Materials Department, Materials Research Laboratory, and California NanoSystems Institute\\
University of California Santa Barbara, California 93106 United States}
\email{yoey@ucsb.edu}

\author{Farnaz Kaboudvand}
\affiliation{Materials Department, Materials Research Laboratory, and California NanoSystems Institute\\
University of California Santa Barbara, California 93106 United States}

\author{Brenden R. Ortiz}
\affiliation{Materials Department, Materials Research Laboratory, and California NanoSystems Institute\\
University of California Santa Barbara, California 93106 United States}

\author{Ram Seshadri}
\affiliation{Materials Department, Materials Research Laboratory, and California NanoSystems Institute\\
University of California Santa Barbara, California 93106 United States}
\email{seshadri@ucsb.edu}

\author{Stephen D. Wilson}
\affiliation{Materials Department, Materials Research Laboratory, and California NanoSystems Institute\\
University of California Santa Barbara, California 93106 United States}
\email{stephendwilson@ucsb.edu}

\date{\today}

\begin{abstract}

The family of \textit{A}V$_3$Sb$_5$ (\textit{A} = K, Rb, Cs) kagome metals exhibit charge density wave (CDW) order, proposed to be chiral, 
followed by a lower temperature superconducting state. Recent studies have proposed the importance of band structure saddle points 
proximal to the Fermi energy in governing these two transitions. Here we show the effects of hole-doping 
achieved \textit{via} chemical substitution of Sn for Sb on the CDW and superconducting states in both KV$_3$Sb$_5$ and 
RbV$_3$Sb$_5$, and generate a phase diagram. Hole-doping lifts the $\Gamma$ pocket and van Hove singularities (vHs) toward 
$E_F$ causing the superconducting $T_C$ in both systems to increase to about 4.5\,K, while rapidly suppressing the 
CDW state. 
\end{abstract}

\maketitle

\section{Introduction}

The \textit{A}V$_3$Sb$_5$ (\textit{A}: K, Rb, Cs) class of kagome
superconductors \cite{ortiz2019new, ortiz2020cs,ortiz2021superconductivity,jiang2021unconventional} are proposed to host both unconventional CDW \cite{feng2021low,park2021electronic,lin2021complex} and SC states \cite{kiesel2013unconventional,wu2021nature,lin2021kagome} and are candidates for realizing these two states intertwined together within a host, pair density wave state \cite{chen2021roton,ge2022discovery,agterberg2011conventional,zhou2021chern}. One means of exploring this coupling is through chemical or mechanical perturbation of the host lattice and the parent electronic state of \textit{A}V$_3$Sb$_5$ compounds.  Recently, the effects of making small 
changes in the electronic band structures on SC and CDW order parameters have been studied 
computationally \cite{labollita2021tuning,uykur2021low} and experimentally by applying external 
hydrostatic pressure \cite{du2021pressure,chen2021double,zhu2022double,yu2021unusual} and hole doping 
on various sites \cite{oey2021fermi,yang2021doping,liu2021doping}.

The electron pocket comprising Sb $p_z$ orbitals at $\Gamma$ and saddle points arising from V $d$-orbitals at 
$M$-points are regarded as playing important roles within the SC and CDW states.  Consequently, resolving their behaviors
when subject to external perturbations such as pressure and chemical doping is important to understand. Recently, 
the effect of hole doping in CsV$_3$Sb$_5$ with Sn substituting Sb was reported, and a double superconducting
dome was observed under light hole-substitution \cite{oey2021fermi}. The termination of the second dome was attributed to the lifting of the $\Gamma$ pocket above $E_F$ and the removal of the Sb $p_z$ orbitals from the Fermi surface; however the formation of the first dome remains anomalous. Within the first SC dome, CDW order rapidly vanishes at doping levels corresponding to only slight movement of the the van Hove singularities (vHs) at the $M$ point toward $E_F$ \cite{oey2021fermi}. 
Curiously, while applied external pressure moves the $M$ saddle points away from $E_F$, it achieves a similar phase diagram suggesting a delicate balance of energy scales within the unperturbed parent structure \cite{labollita2021tuning, chen2021double} and the dominance of removing the CDW state in governing the formation of the first SC dome.

In contrast to the above, pressure studies of KV$_3$Sb$_5$ and RbV$_3$Sb$_5$ have reported only a single superconducting dome coinciding with the suppression of CDW order  \cite{du2021pressure,zhu2022double}; however studies of carrier-induced perturbations of these variants remain unexplored.  To further explore similarities between pressure-induced and carrier-induced perturbations of the interplay between CDW and SC orders, here the effect of hole doping \textit{A}V$_3$Sb$_5$ with in solid solutions with \textit{A} = K, Rb is presented. 
In both systems, only one superconducting dome is observed in contrast with CsV$_3$Sb$_{5-x}$Sn$_x$, although a similar, rapid CDW suppression is observed. The solubility limit of Sn in KV$_3$Sb$_5$ is the lowest in the family, with phase separation observed by $x = 0.30$. In contrast RbV$_3$Sb$_5$ supports up to $x = 0.70$ of Sn replacing Sb. While the differences in solubility can be explained by \textit{A} cation size effects, the difference between the single dome hole-doping phase diagrams of KV$_3$Sb$_5$ and RbV$_3$Sb$_5$ relative to double dome phase diagram of CsV$_3$Sb$_5$ suggests a difference in the parent CDW state between the three materials. The potential origins of these differences are discussed.

\section{Experimental details}

\begin{figure*}
\includegraphics[width=1\textwidth]{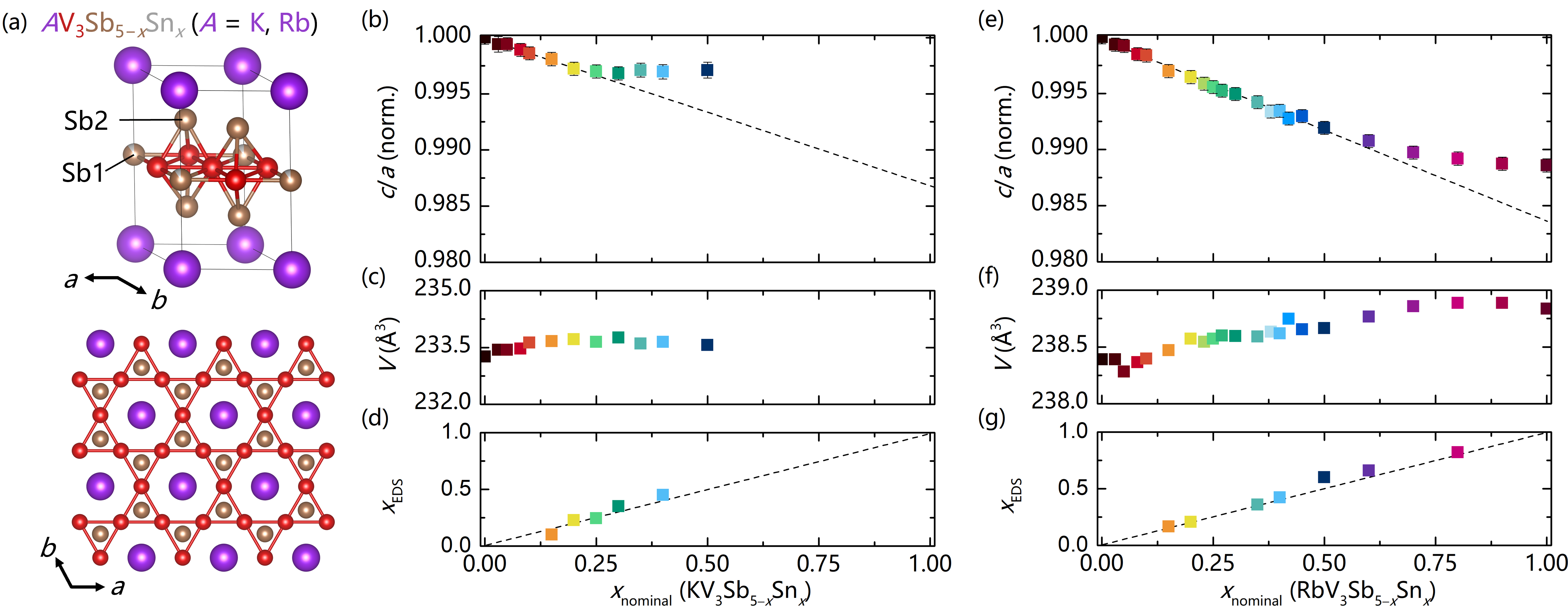}
\caption{(a) \textit{A}V$_3$Sb$_{5-x}$Sn$_x$ crystallizes in the parent \textit{A}V$_3$Sb$_5$ ($P$6/$mmm$) structure. 
Naively Sn can substitute on either the Sb1 or Sb2 sites, but from studies on CsV$_3$Sb$_{5-x}$Sn$_x$ it is clear 
that Sn preferentially substitutes in the in-plane kagome Sb1. (b-c) Sn integration into KV$_3$Sb$_{5-x}$Sn$_x$ 
causes the $c/a$ ratio to decrease, and the solubility limit of Sn is $x \approx$ 0.25, at which point the $c/a$ 
ratio deviates from the linear trend. (e-f) The solubility limit of Sn in RbV$_3$Sb$_{5-x}$Sn$_x$ is a bit higher, 
at $x \approx$ 0.70. (d, g) For samples with concentrations above the EDS sensitivity threshold, nominal Sn and 
measured Sn content agree. Error bars are shown unless they are smaller than associated point size.}
\label{fig:1structure}
\end{figure*}

Powder samples of KV$_3$Sb$_{5-x}$Sn$_x$ and RbV$_3$Sb$_{5-x}$Sn$_x$ for 0 $\leq x \leq$ 1 were synthesized as 
described previously \cite{oey2021fermi}. Structural analysis was performed \textit{via} a Panalytical Empyrean laboratory x-ray powder diffractometer. A Hitachi TM4000Plus scanning electron microscope (SEM) was used to perform energy--dispersive X--ray spectroscopy (EDS). A Quantum Design Magnetic Property Measurement System (MPMS) was used to collect magnetization data and a Quantum Design Physical Property Measurement System (PPMS) Dynacool was used for resistivity measurements. To gain insight into the electronic states of all compounds, first-principles calculations based on density functional theory (DFT) within the Vienna ab initio Simulation Package (VASP) were performed \cite{kresse1996efficient,kresse1996efficiency}.

\begin{figure*}
\includegraphics[width=1\textwidth]{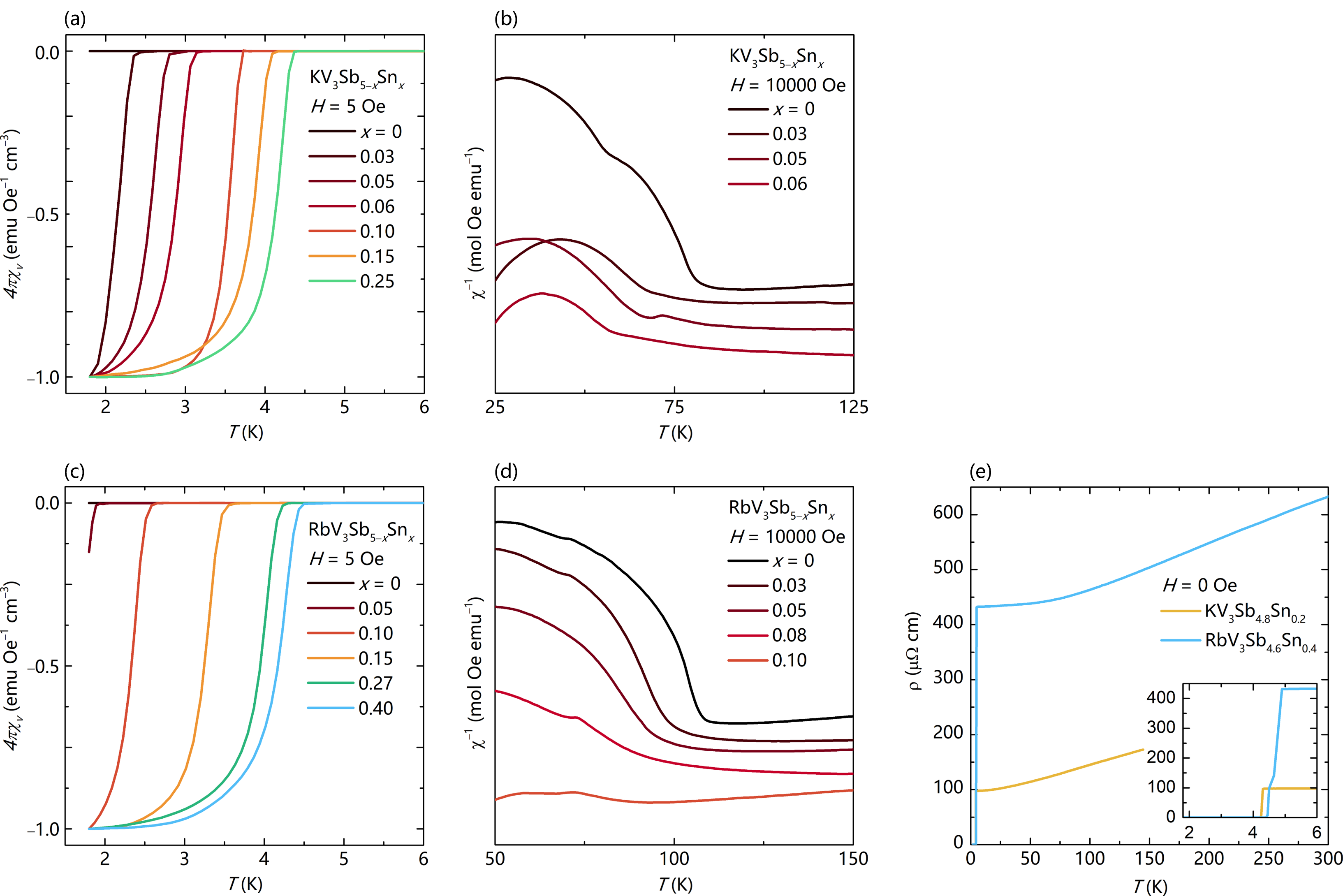}
\caption{(a) The superconducting $T_C$ of KV$_3$Sb$_{5-x}$Sn$_x$ measured under a field of $H$ = 5\,Oe systematically shifts to higher temperature in compositions up to $x = 0.30$. The superconducting fraction is normalized to account for errors in mass and packing fraction so all data have a minimum of $-$1, the theoretical minimum. (b) $1/\chi$ for compositions $x \leq$ 0.05 in KV$_3$Sb$_{5-x}$Sn$_x$ show the CDW $T^*$ decreases from 86\,K to 59\,K and disappears for $x \geq 0.05$. (c) The superconducting $T_C$ of RbV$_3$Sb$_{5-x}$Sn$_x$ measured under a field of $H$ = 5\,Oe systematically shifts to higher temperature in compositions up to $x = 0.40$. (d) $1/\chi$ for compositions $x \leq$ 0.10 in RbV$_3$Sb$_{5-x}$Sn$_x$ show the CDW $T^*$ decreases and eventually disappears for $x \geq 0.10$. (e) Resistivity data for KV$_3$Sb$_{4.8}$Sn$_{0.2}$ and RbV$_3$Sb$_{4.6}$Sn$_{0.4}$ confirm the $T_C$ transitions and lack of CDW transitions for these higher Sn content samples.}
\label{fig:2SCCDWro}
\end{figure*}

\section{Experimental and Computational Results}

\textit{A}V$_3$Sb$_{5-x}$Sn$_x$ all adopt the parent hexagonal $P6/mmm$ structure at room temperature, with the V atoms making up an ideal kagome network. $^{121}$Sb nuclear quadrupole resonance (NQR) studies on CsV$_3$Sb$_{5-x}$Sn$_x$ 
have shown that the Sn atoms preferentially occupy the Sb1 sublattice site in the kagome plane \cite{oey2021fermi}. 
Due to the similarity in structure in all of the \textit{A}V$_3$Sb$_5$ compounds, we assume that the Sn atoms occupy 
the Sb1 sublattice for \textit{A} = K, Rb as well. For KV$_3$Sb$_{5-x}$Sn$_x$, polycrystalline samples were found 
to be single phase for $x \leq 0.25$, while for RbV$_3$Sb$_{5-x}$Sn$_x$, polycrystalline samples were single phase 
for $x \leq 0.7$, at which point the limit of the solid solution was reached and secondary phases were observed in 
both families.

Powder diffraction data were fitted using the Pawley method to study changes in unit cell as a function of Sn content. Similar to the changes observed in CsV$_3$Sb$_{5-x}$Sn$_x$, $a$ increases as $c$ decreases with increasing Sn content in both \textit{A} = K, Rb as seen in Fig.\,\ref{fig:1structure}(b,e), while the volume is independent of Sn content (Fig.\,\ref{fig:1structure}(c,f)). The solubility limit of Sn is clearly reached once the $c/a$ ratio deviates from its linear trend ($x \approx 0.30$ for KV$_3$Sb$_{5-x}$Sn$_x$ and $x \approx 0.70$ for RbV$_3$Sb$_{5-x}$Sn$_x$), and a secondary phase emerges in the powder diffraction data. EDS was performed on samples with a critical Sn content to confirm the nominal Sn content (Fig.\,\ref{fig:1structure}(d, g)).

Single crystal KV$_3$Sb$_5$ has a superconducting $T_C$ of 0.93\,K\cite{ortiz2021superconductivity}. Here, the $T_C$ of polycrystalline KV$_3$Sb$_5$ is lower than 1.8\,K, the lowest achievable temperature on the MPMS, but as Sn was incorporated, the superconducting transition temperature increases and can be detected by $x = 0.03$. Figure\,\ref{fig:2SCCDWro}(a) shows the evolution of the superconducting transition up to $x = 0.25$, where $T_C$ is at a maximum of 4.5\,K. Beyond this Sn content, a secondary phase appears, so an eventual suppression of superconductivity is not observed within the solid solution of KV$_3$Sb$_{5-x}$Sn$_x$. RbV$_3$Sb$_5$ shows similar behavior, with undoped crystals showing a $T_C$ of 0.92\,K\cite{yin2021superconductivity}. Again, although the superconducting transition for RbV$_3$Sb$_5$ cannot be detected with an MPMS, by $x = 0.05$ the transition appears at 1.93\,K and increases to a maximum of 4.5\,K for $x = 0.40$, as seen in Figure\,\ref{fig:2SCCDWro}(c). For $x > 0.40$, $T_C$ decreases to 3.55\,K until a clear secondary phase appears. All of the superconducting volume fractions are approximately $4 \pi \chi_v \approx -1$ and are normalized to $-1$ for ease of comparison; fractions that deviate slightly from this ideal value can be attributed to errors in packing density.

The CDW $T^*$ was determined by looking at the inflection point of $\chi^{-1}$ \textit{vs.} $T$. Polycrystalline KV$_3$Sb$_5$ as observed here has a $T^*$ of 86\,K, which is in good agreement with reported CDW transition temperatures of $\approx$ 80\,K in crystals \cite{jiang2021unconventional}. Using this metric, the CDW transition is quickly suppressed with increasing hole-doping, decreases to 60\,K by $x = 0.06$, and is fully suppressed by $x = 0.06$. In RbV$_3$Sb$_5$, the CDW $T^*$ using the same metric is $\approx$ 112\,K, similar to the literature reported value \cite{ortiz2019new}. The CDW is quickly suppressed to 91\,K for $x = 0.10$ at which point it fully disappears (Fig.\,\ref{fig:2SCCDWro}(d)). In both systems, the CDW transition is fully suppressed with much less Sn content than that required to reach the peak of the superconductivity dome, suggesting that the interplay between superconductivity and CDW is different in \textit{A} = K, Rb than in CsV$_3$Sb$_5$, where the CDW persists through the peak of the first superconducting dome. 

Electrical resistivity measurements of samples near the superconducting peak concentrations are shown in Figure\,\ref{fig:2SCCDWro}(e). The superconducting states are clearly observed as zero-resistivity conditions and the transitions at 4.25\,K for KV$_3$Sb$_{4.80}$Sn$_{0.20}$ and 4.4\,K for RbV$_3$Sb$_{4.60}$Sn$_{0.40}$ agree well with those seen in the magnetization data, while signatures of CDW transitions are also not present. Consistent with CsV$_3$Sb$_{5-x}$Sn$_{x}$, the residual resistance increases in RbV$_3$Sb$_{5-x}$Sn$_{x}$ and KV$_3$Sb$_{5-x}$Sn$_{x}$ under Sn-substitution while $T_C$ increases.

\begin{figure}
\includegraphics[width=0.9\linewidth]{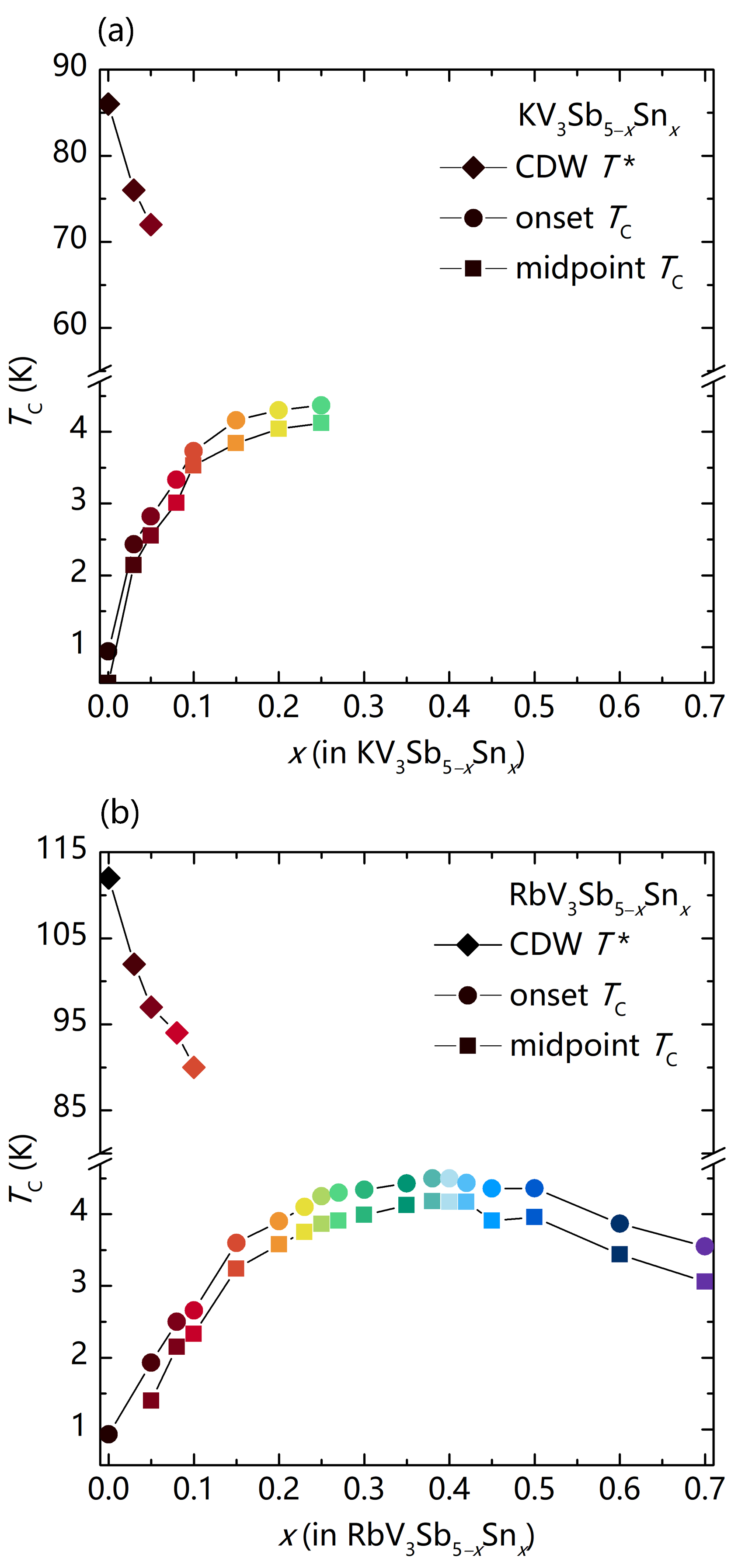}
\caption{The progression of superconducting $T_C$ and CDW order $T^*$ is plotted as a function of Sn composition for KV$_3$Sb$_{5-x}$Sn$_x$ and RbV$_3$Sb$_{5-x}$Sn$_x$. In KV$_3$Sb$_{5-x}$Sn$_x$, the solubility limit of Sn is reached before a full superconducting dome is achieved, but the suppression of CDW $T^*$ is still fully realized. RbV$_3$Sb$_{5-x}$Sn$_x$ shows a single superconducting dome before the solubility limit is reached, with an accompanying suppression of CDW order by $x = 0.1$.}
\label{fig:3phasediagram}
\end{figure}

The effects of Sn-substitution on SC and CDW orders in KV$_3$Sb$_{5-x}$Sn$_x$ and RbV$_3$Sb$_{5-x}$Sn$_x$ are summarized in Figure \ref{fig:3phasediagram}. In both systems, the CDW ordering temperature is rapidly suppressed with small levels of hole doping and fully disappears before the maximum superconducting transition temperatures.  With the exception of the soluability limit of Sn in each lattice, the two phase diagrams are qualitatively similar.

\begin{figure*}
\includegraphics[width=1\textwidth]{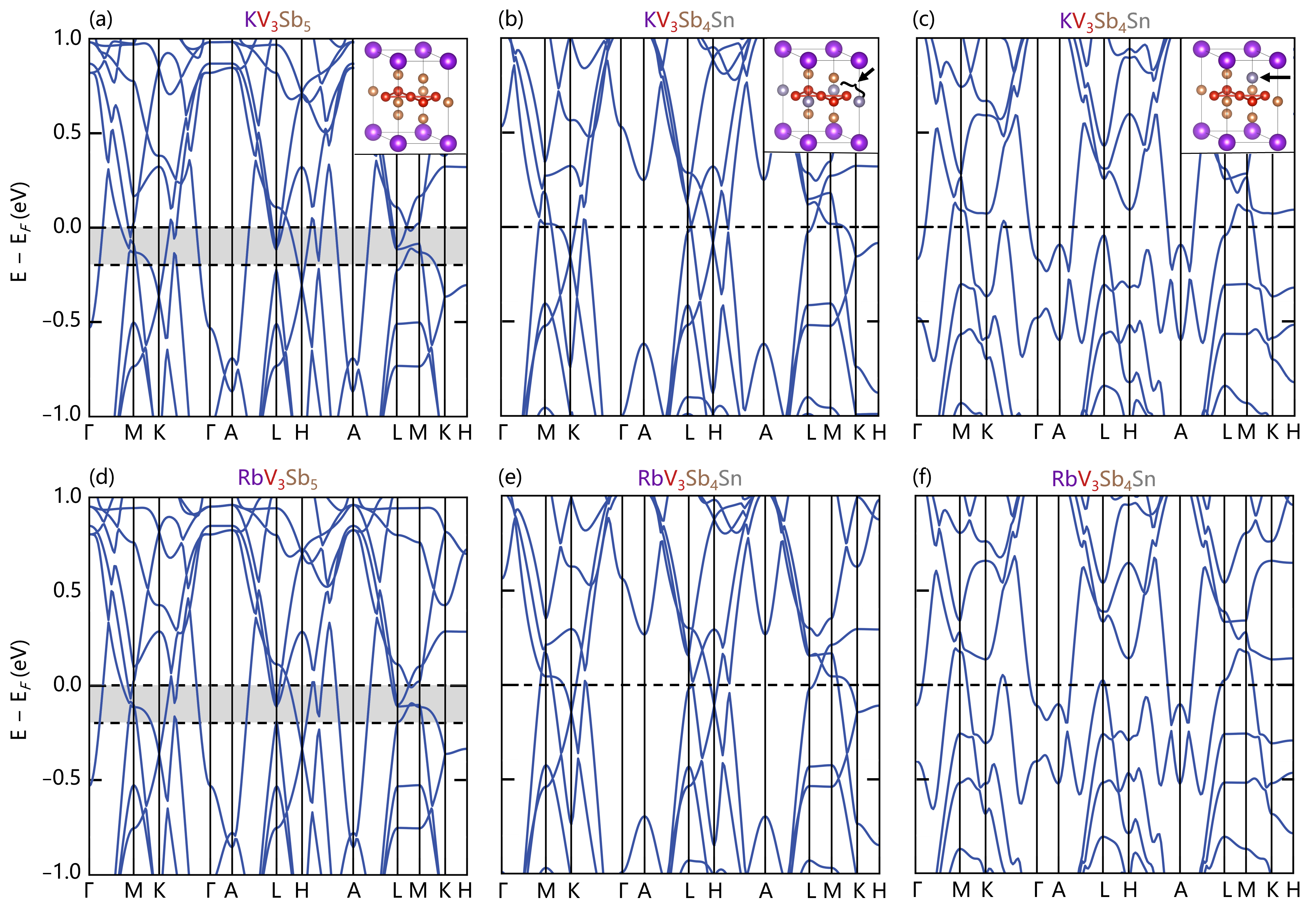}
\caption{(a,d) DFT calculations for KV$_3$Sb$_5$ and RbV$_3$Sb$_5$ highlighting the allowable range of Fermi levels under the rigid band approximation for substitution of one Sn atom per formula unit (one electron less Fermi energy). (b,e) Calculation for a hypothetical structure where one Sn has been substituted within the kagome plan (Sb1 sublattice).  The majority of the electronic structure is preserved as KV$_3$Sb$_5$ and RbV$_3$Sb$_5$, except the $\Gamma$ pocket and M point van Hove singularities which have been shifted far above and may contribute to the changing observable properties. (c,f) Calculation for a hypothetical structure where one Sn has been substituted out of the kagome plane (Sb2 sublattice). Here a strong reconstruction of many bands can be observed, in particular near K, H, and K-$\Gamma$.}
\label{fig:4dft}
\end{figure*}

Due to size considerations, the solubility limits of Sn in KV$_3$Sb$_5$ and RbV$_3$Sb$_5$ are much lower than in CsV$_3$Sb$_5$. However in all of these systems, the changes in electronic structure for the compositions of interest can be extrapolated by computing band structures of substitution one Sb atom for an Sn atom within the kagome plane and out of the kagome plane. Figure\,\ref{fig:4dft}(a) and (d) shows the band structure of undoped KV$_3$Sb$_5$ and RbV$_3$Sb$_5$, respectively, with the range of achievable Fermi levels with the loss of 1 electron per unit cell indicated in gray. Figure \ref{fig:4dft}(b) and (e) show a hypothetical structure where the Sb1 lattice is fully substituted for Sn. The shifts in band structure are similar to those seen in CsV$_3$Sb$_4$Sn, with the $\Gamma$ pocket and $M$-point vHs lifted above the Fermi level. Band structures for Sn substituted within the Sb2 sublattice are plotted in Figure \ref{fig:4dft}(c) and (f) and show significant reconstructions at multiple points, including $K$, $L$, and $H$.

\section{Discussion}

Given the close structural and electronic properties across the $A$V$_3$Sb$_5$ series, the difference in the hole-doping phase diagrams of KV$_3$Sb$_5$ and RbV$_3$Sb$_5$ relative to CsV$_3$Sb$_5$ is anomalous. The changes in band structure for Sn substitution on the Sb1 sublattice in KV$_3$Sb$_4$Sn and RbV$_3$Sb$_4$Sn at first sight, appear to be very similar to those in CsV$_3$Sb$_4$Sn. However, subtle differences are nevertheless present and potentially lead to distinct electronic phase diagrams. A closer examination of the saddle points around $M$ reveals that the irreducible representations of the two points are not the same for \textit{A} = K, Rb and Cs \cite{labollita2021tuning,uykur2021low}. As a consequence of the saddle point inversions, bands cross near $E_F$ in CsV$_3$Sb$_5$ but not in KV$_3$Sb$_5$. This may play a role in the nature of the CDW states stabilized, whose proximity to one another may modify SC in unique ways. 

One unifying theme amongst doping and pressure studies of $A$V$_3$Sb$_5$ is the similarity of two SC domes present uniquely in CsV$_3$Sb$_5$.  If one envisions either pressure/doping empirically as a perturbation that simply destabilizes CDW order, then the data suggest that CsV$_3$Sb$_5$ hosts a CDW state distinct from the other two variants.  In this scenario, there is a transition between CDW states within the phase diagram of CsV$_3$Sb$_5$ that is naively absent in KV$_3$Sb$_5$ and RbV$_3$Sb$_5$.  

While the details of the intermediate states differ between pressure and hole-doping studies, investigations of the parent CDW order in $A$V$_3$Sb$_5$ have shown differences between $A$=Cs and $A=$Rb/K.  The out-of-plane modulation for Cs is four times that of K \cite{ortiz2021fermi,jiang2021unconventional}, and the in-plane structure of the CDW state in CsV$_3$Sb$_5$ hosts both star-of-David and tri-hexagonal character \cite{ortiz2021fermi,hu2022coexistence,kang2022microscopic}, a pattern seemingly not present in the other variants. Given these differences, one possibility is a doping-induced transition from an ($L$, $L$, $L$)-type order to an ($M$, $L$, $L$)-type of order upon doping CsV$_3$Sb$_5$ \cite{christensen2021theory} that is not present in the other two compounds.  Future scattering studies across the carrier-tuned phase diagrams of these systems will be required to fully explore any potential crossover.

\section{Conclusion}
 The Sn-substitution or hole-doping phase diagrams of KV$_3$Sb$_5$ and RbV$_3$Sb$_5$ were explored.  The solubility of Sn within the $A$V$_3$Sb$_5$ lattice varies with $A$-site cation size with $A$=K having the smallest solubility and $A$=Cs having the largest.  Prior to reaching the solubility limit, hole-doping achieved \textit{via} substitution of Sn on Sb sites in KV$_3$Sb$_5$ and RbV$_3$Sb$_5$ reveals similar phase diagrams consisting of a single SC dome and rapidly suppressed CDW order. This contrasts the recently reported double dome observed in CsV$_3$Sb$_5$ and suggests a distinct parent CDW state for the latter.

\section{acknowledgments}
This work was supported by the National Science Foundation (NSF) through Enabling Quantum Leap: Convergent Accelerated Discovery Foundries for Quantum Materials Science, Engineering and Information (Q-AMASE-i): Quantum Foundry at UC Santa Barbara (DMR-1906325). The research reported here made use of shared facilities of the NSF Materials Research Science and Engineering Center at UC Santa Barbara DMR-1720256, a member of the Materials Research Facilities Network (www.mrfn.org). Use of the Advanced Photon Source at Argonne National Laboratory was supported by the U.S. Department of Energy, Office of Science, Office of Basic Energy Sciences, under Contract No. DE-AC02-06CH11357. YMO is supported by the National Science Foundation Graduate Research Fellowship Program under Grant No. DGE-1650114. FK acknowledges the Roy T. Eddleman Center for Quantum Innovation (ECQI) for their support.

\section*{References}

\bibliography{135frustratedmagnets}

\end{document}